%% file: main.tex
\documentclass[sigconf]{acmart}
\renewcommand\footnotetextcopyrightpermission[1]{}

\usepackage{textcomp}
\usepackage{verbatim}
\usepackage{color}
\usepackage{xcolor, colortbl}
\definecolor{graytwo}{gray}{.7}
\usepackage{xspace}
\usepackage{balance}
\usepackage{tabulary}
\usepackage{array} 
\usepackage[ruled,lined,linesnumbered,vlined,algo2e]{algorithm2e}
\usepackage{amsmath}
\usepackage{subcaption}
\usepackage{listings}
\usepackage{lipsum}
\usepackage{hyperref}
\usepackage{cleveref}
\usepackage{wrapfig}
\usepackage{graphicx}
\usepackage[T1]{fontenc}
\usepackage[ansinew]{inputenc}
\usepackage[many]{tcolorbox}
\usepackage{extarrows}
\usepackage{booktabs}
\usepackage{tabularx}
\usepackage{enumitem}
\setitemize[0]{leftmargin=15pt}
\setlist{nosep}

\usepackage{xurl}

\usepackage[justification=centering]{caption}
\usepackage{pifont}
\usepackage{makecell,multirow,diagbox,rotating}
\usepackage{longtable}
\usepackage{etoolbox} 
\usepackage{algorithm}
\usepackage{algorithmicx} 
\usepackage{algpseudocode}
\usepackage{amsmath}
\usepackage{amsfonts}

\usepackage{amssymb}
\usepackage{dutchcal}
\usepackage{breakurl}
\newcommand{\hlt}[1]{\textbf{\textcolor{red!70!black}{#1}}}
\definecolor{codegreen}{rgb}{0,0.6,0}
\definecolor{codegray}{rgb}{0.5,0.5,0.5}
\definecolor{codepurple}{rgb}{0.58,0,0.82}
\definecolor{backcolour}{rgb}{0.95,0.95,0.92}

\lstdefinelanguage{json}{
  string=[s]{"}{"},
  stringstyle=\color{blue}\ttfamily,
  comment=[l]{//},
  commentstyle=\color{codegray}\ttfamily,
  literate=
    *{:}{{{\color{violet}{:}}}}{1}
    {,}{{{\color{violet}{,}}}}{1}
}

\lstdefinelanguage{Swift}{
  keywords={class, func, var, let, if, else, guard, return, import, struct, enum, protocol, extension, private, public, async, await, throws, throw, try, catch, for, in, while, do, switch, case, break, continue, default, static, self, Self, true, false, nil, some, where, associatedtype, typealias, init, deinit, subscript, override, convenience, required, lazy, weak, unowned, willSet, didSet, inout, operator, prefix, postfix, precedencegroup},
  keywordstyle=\color{violet}\bfseries,
  ndkeywords={String, Int, Double, Bool, Array, Dictionary, Set, Optional, Any, AnyObject, URL, URLRequest, Data, Error, Result, Task, MainActor},
  ndkeywordstyle=\color{darkgray}\bfseries,
  identifierstyle=\color{black},
  sensitive=true,
  comment=[l]{//},
  morecomment=[s]{/*}{*/},
  commentstyle=\color{codegray}\ttfamily,
  stringstyle=\color{blue}\ttfamily,
  morestring=[b]",
  frame=none,
  numbers=none
}

\lstdefinelanguage{JavaScript}{
  keywords={typeof, new, true, false, catch, function, return, null, catch, switch, var, if, in, while, do, else, case, break, const, prototype},
  keywordstyle=\color{violet}\bfseries,
  ndkeywords={class, export, boolean, throw, implements, import, this},
  ndkeywordstyle=\color{darkgray}\bfseries,
  identifierstyle=\color{black},
  sensitive=false,
  comment=[l]{//},
  morecomment=[s]{/*}{*/},
  commentstyle=\color{gray}\ttfamily,
  stringstyle=\color{blue}\ttfamily,
  morestring=[b]',
  morestring=[b]",
  frame=none,
  numbers=none
}

\usepackage{fancybox}


\makeatletter
\renewcommand{\maketag@@@}[1]{\hbox{\m@th\normalsize\normalfont#1}}%
\makeatother

\usepackage{makecell}
\usepackage{threeparttable}
\makeatletter  
\newif\if@restonecol  
\renewcommand\footnoterule{%
	\kern-3\p@
	\hrule\@width\columnwidth
	\kern2.6\p@}

\usepackage{url}

\SetCommentSty{mycommfont}

\definecolor{Green}{RGB}{0,180,0}
\newcommand{\tool}{\textsc{LLMKeyLens}\xspace}
\newcommand{\totalCollectApp}{38,520\xspace}
\newcommand{\totaldupApp}{5,619\xspace}

\newcommand{\totalevaApp}{1092\xspace}

\newcounter{findingcounter}
\newcommand{\finding}[1]{%
\refstepcounter{findingcounter}%
\vspace{0.1em}
\begin{tcolorbox}
[colback=gray!8, size=title, boxsep=1mm, colframe=black,after={\vskip0mm}]
\textbf{Finding \thefindingcounter:} \emph{#1}
\end{tcolorbox}
}

\newcolumntype{L}[1]{>{\raggedright\let\newline\\\arraybackslash\hspace{0pt}}m{#1}}
\newcolumntype{C}[1]{>{\centering\let\newline\\\arraybackslash\hspace{0pt}}m{#1}}
\newcolumntype{R}[1]{>{\raggedleft\let\newline\\\arraybackslash\hspace{0pt}}m{#1}}

\AtBeginDocument{%
  \providecommand\BibTeX{{%
    \normalfont B\kern-0.5em{\scshape i\kern-0.25em b}\kern-0.8em\TeX}}}

\input{stats}

\begin{document}

\title{ Mind your key: An Empirical Study of LLM API Credential Leakage in iOS Apps}

\settopmatter{authorsperrow=2}  

\author{Pinran Gao}
\authornote{Equal contribution.}
\affiliation{%
  \institution{Wake Forest University}
  \city{Winston-Salem}
  \state{North Carolina}
  \country{USA}
}
\email{gaop24@wfu.edu}

\author{Wang Lingxiang}
\authornotemark[1]
\affiliation{%
  \institution{Unaffiliated}
  \country{}
}
\email{lingxiang.wang.2016@gmail.com}

\author{Yi Liu}
\affiliation{%
  \institution{Griffith University}
  \country{}
}
\email{yi009@e.ntu.edu.sg}

\author{Kunpeng Liu}
\affiliation{%
  \institution{Clemson University}
  \country{}
}
\email{kunpenl@clemson.edu}

\author{Fan Yang}
\affiliation{%
  \institution{Wake Forest University}
  \city{Winston-Salem}
  \state{North Carolina}
  \country{USA}
}
\email{yangfan@wfu.edu}

\author{Ying Zhang}
\affiliation{%
  \institution{Wake Forest University}
  \city{Winston-Salem}
  \state{North Carolina}
  \country{USA}
}
\email{ying.zhang@wfu.edu}

\input{abstract}


\maketitle

\input{intro}

\input{background}

\input{methodlogy/overview}

\input{experiment/expsetting}

\input{related}
\input{Discussion}
\input{Threats}

\input{Conclusion}


\balance
\bibliographystyle{ACM-Reference-Format}
\bibliography{reference}

\end{document}

%% file: stats.tex
\newcommand{\totalApps}{1092}           
\newcommand{\totalTestApps}{444}        
\newcommand{\leakedTotal}{282}
\newcommand{\leakedPct}{26\%}           
\newcommand{\leakedTestPct}{64\%}       
\newcommand{\leakedFull}{146}
\newcommand{\leakedFullPct}{52\%}
\newcommand{\leakedLimited}{136}
\newcommand{\noLeak}{162}

\newcommand{\ratingMedian}{16}           
\newcommand{\ratingMean}{12,851}         
\newcommand{\ratingOverKPct}{15\%}       
\newcommand{\ratingMax}{2,305,613}       

\newcommand{\leakedCategories}{13}
\newcommand{\prodLeaked}{142}
\newcommand{\prodLeakedPct}{50\%}
\newcommand{\prodTotal}{355}
\newcommand{\prodLeakRate}{40\%}

\newcommand{\entLeaked}{34}
\newcommand{\entLeakedPct}{12\%}
\newcommand{\lifeLeaked}{24}
\newcommand{\lifeLeakedPct}{9\%}
\newcommand{\healthLeaked}{7}
\newcommand{\healthTotal}{15}
\newcommand{\healthLeakRate}{47\%}
\newcommand{\refLeaked}{4}
\newcommand{\refTotal}{13}
\newcommand{\refLeakRate}{31\%}
\newcommand{\finTotal}{5}
\newcommand{\medTotal}{10}

\newcommand{\directApps}{60}
\newcommand{\directPct}{21\%}
\newcommand{\cloudApps}{67}
\newcommand{\cloudPct}{23\%}
\newcommand{\customApps}{155}
\newcommand{\customPct}{55\%}
\newcommand{\openaiApps}{42}
\newcommand{\openaiPct}{15\%}
\newcommand{\geminiApps}{7}

\newcommand{\firebaseApps}{27}

\newcommand{\gcrApps}{11}

\newcommand{\directKeyApps}{54}
\newcommand{\directKeyPct}{19\%}

\newcommand{\jwtApps}{136}
\newcommand{\jwtPct}{48\%}

\newcommand{\noAuthApps}{92}
\newcommand{\noAuthPct}{33\%}

\newcommand{\analyzableApps}{444}
\newcommand{\mechApps}{143}
\newcommand{\mechPct}{32\%}

\newcommand{\totalBypassed}{58}
\newcommand{\totalBypassRate}{41\%}
\newcommand{\totalResisted}{85}
\newcommand{\totalResistedPct}{59\%}
\newcommand{\singleApps}{100}

\newcommand{\singleBypassRate}{55\%}
\newcommand{\encryptBypassed}{2}
\newcommand{\encryptRate}{15\%}
\newcommand{\proxyBypassApps}{59}
\newcommand{\proxyBypassBypassed}{48}
\newcommand{\proxyBypassRate}{81\%}

\newcommand{\netFailApps}{11}
\newcommand{\netFailBypassed}{5}
\newcommand{\netFailRate}{45\%}

\newcommand{\packetFailApps}{8}
\newcommand{\encryptApps}{13}
\newcommand{\antiDetectApps}{5}
\newcommand{\wsApps}{4}
\newcommand{\multiApps}{43}

\newcommand{\multiBypassRate}{7\%}
\newcommand{\multiReduction}{7.9}
\newcommand{\proxyEncApps}{15}
\newcommand{\proxyEncBypassed}{0}
\newcommand{\proxyEncRate}{0\%}
\newcommand{\wsProxyApps}{13}
\newcommand{\wsProxyBypassed}{1}
\newcommand{\wsProxyRate}{8\%}
\newcommand{\proxyNetFailApps}{5}
\newcommand{\proxyNetFailBypassed}{1}
\newcommand{\proxyNetFailRate}{20\%}
\newcommand{\wsProxyEncApps}{4}
\newcommand{\wsEncApps}{2}
\newcommand{\multiOtherApps}{4}
\newcommand{\multiOtherBypassed}{1}
\newcommand{\multiOtherRate}{25\%}
\newcommand{\proxyInMultiPct}{93\%}

\newcommand{\testedExploitRate}{64\%}

\newcommand{\retestTotal}{282}
\newcommand{\retestFull}{146}
\newcommand{\retestLimited}{136}
\newcommand{\retestStillExploitable}{66}
\newcommand{\retestStillExploitablePct}{23\%}
\newcommand{\retestRevoked}{78}
\newcommand{\retestRevokedPct}{28\%}
\newcommand{\retestRateLimited}{9}

\newcommand{\retestUnreachable}{83}
\newcommand{\retestUnreachablePct}{29\%}

\newcommand{\retestYesStill}{36}

\newcommand{\retestYesNoAuth}{31}
\newcommand{\retestYesDirectKey}{5}

\newcommand{\retestYesRevoked}{27}
\newcommand{\retestLimitStill}{30}
\newcommand{\retestLimitStillPct}{22\%}
\newcommand{\retestLimitRevoked}{51}
\newcommand{\retestLimitNoExp}{5}
\newcommand{\retestLimitExpiredAccepted}{5}
\newcommand{\retestLimitExcessive}{6}
\newcommand{\retestLimitStatic}{13}

\newcommand{\retestLimitSessionToken}{1}

%% file: abstract.tex
\begin{abstract}
The rapid integration of large language models (LLMs) into mobile applications has introduced a new class of credential security risk: leaked credentials that grant unauthorized access to LLM inference services, which can cause financial damage to the developer side. Prior work has studied credential leakage across various platforms, with a primary focus on Android Apps. However, to date, no empirical study has systematically investigated how LLM API key leakage occurs in iOS applications.

In this paper, we conducted the first in-depth empirical study of API key leakage in LLM-integrated applications. We constructed a high-quality dataset of \totalTestApps{} iOS applications by filtering from \totalApps{} candidate apps through a standardized evaluation process. 
To capture key leakage, we developed \tool, a dynamic analysis framework that detects LLM API key leakage in iOS applications. \tool automates traffic interception, provider-specific key extraction, and active validity confirmation, requiring neither source code access nor binary decryption.
We apply \tool to \totalTestApps{} applications. 
Our analysis reveals that \textbf{\leakedTotal{}} applications expose exploitable LLM API credentials in network traffic, spanning at least ten LLM providers. We identify three distinct leakage patterns.
The most prevalent is JWT-based token leakage  (\jwtPct{}), followed by unauthenticated backend proxy access (\noAuthPct{}) and plaintext API key transmission (\directKeyPct{}). 
To assess remediation effectiveness, we re-analyzed the same \retestTotal{} vulnerable applications three months after 
responsible disclosure. Comparing the two scans, only 
\retestRevokedPct{} of applications had remediated the reported vulnerability, while 72\% remained exploitable. The persistent vulnerabilities stem from unauthenticated backends and broken JWT implementations. 

Our findings reveal that LLM API key leakage is both prevalent and persistent in the iOS ecosystem, exposing a systemic gap between real-world developer practice and secure integration principles. 
With only \retestRevokedPct{} of vulnerable applications remediated after disclosure, our results suggest that secure LLM integration requires not only developer awareness but also explicit security guidance from providers and platform-level enforcement mechanisms.
\end{abstract}

%% file: intro.tex
\section{Introduction}

Many LLM providers, including OpenAI~\cite{openai_api} and Google Gemini~\cite{gemini_api}, as well as an expanding ecosystem of third-party services (e.g., OpenRouter~\cite{openrouter}), expose cloud-hosted inference APIs that enable developers to embed LLM capabilities into their Applications (Apps)~\cite{ozkaya2024llmproviders}. Developers integrate LLM services using provider-issued credentials that authenticate requests and authorize billable usage. As of 2025, LLM-powered apps have reached 17 billion downloads, accounting for 13\% of all mobile app downloads~\cite{sensortower2025}.

\begin{figure}[htbp]
    \centering
    \vspace{-0.5em}
    \includegraphics[width=0.80\linewidth]{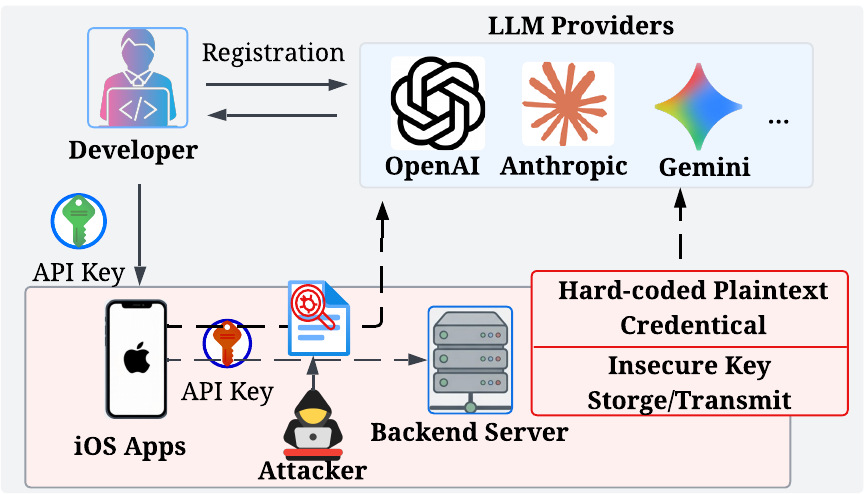}
     \vspace{-0.5em}
    \caption{LLM API credential leakage via network traffic interception. The App sends a request to the LLM API provider or backend proxy with the API key included.}
    \label{fig:keyleakage}
     \vspace{-0.5em}
\end{figure}

Despite the sensitivity of these credentials, many developers manage LLM API keys insecurely~\cite{ibrahim2025lmscout}. They may simply embed these keys directly in client-side App code (see Figure~\ref{fig:keyleakage}), which can cause critical leakage in mobile Apps. Attackers who obtain these leaked keys can launch unauthorized LLM inference requests, thereby shifting the associated financial burden to legitimate App developers. As documented in the Sysdig report~\cite {brucato2024llmjacking}, attackers exploited a web App and stole its LLM API keys, resulting in potential daily financial losses exceeding \$46,000 for the affected organization and widespread service abuse.

\vspace{-0.3em}


Prior research has extensively studied credential leakage in Android Apps~\cite{zhouDevelopercredentialsAndroid2015, zuoCloudApp2019, zhang2023don, weiAppSecretAndroid2025, MendozawebtoMobile2018, LeakyAppSchmidt2025}. For instance, Zhou et al.~\cite{zhouDevelopercredentialsAndroid2015} found over 1,000 Android Apps embedding plaintext cloud service credentials, Wei et al.~\cite{weiAppSecretAndroid2025} empirically analyzed the full attack chain from secret embedding to exploitation, and most recently, Ibrahim et al.~\cite{ibrahim2025lmscout} examined 2,600 Android Apps and found LLM-specific API keys hardcoded in App binaries. However, the iOS ecosystem presents unique challenges that have left it largely unstudied. Unlike Android, where APK files can be readily decompiled, iOS Apps distributed through the Apple App Store are protected by FairPlay DRM~\cite{wikipedia_fairplay}, which makes static binary scanning infeasible without jailbreaking and decryption. Moreover, even approaches that successfully decrypt iOS binaries (e.g., Schmidt et al.~\cite{LeakyAppSchmidt2025}) require a jailbroken device as a prerequisite and may produce false positives from inactive or obfuscated credentials.

In this work, we conduct the first in-depth analysis of LLM API credential leakage in iOS Apps. We first constructed a dataset of \totaldupApp{} LLM-related unique iOS Apps collected from the US App Store using iTunes search APIs~\cite{apple_search_api} in Oct 2025. Since the App Store requires authenticated sessions for each download, batch automated retrieval is infeasible; we therefore sampled \textbf{\totalevaApp{}} Apps for manual evaluation. 
Six trained evaluators, organized into three pairs, independently screened each App on physical iOS devices following standardized exclusion criteria, removing Apps that could not be downloaded, launched, or exercised due to upfront purchase requirements, registration barriers, or absence of observable LLM functionality. This process yielded a high-quality dataset of \textbf{\totalTestApps{}} Apps with testable LLM features. 

To detect credential leakage without requiring jailbreaking or binary decryption, we further developed \tool, a dynamic analysis framework built based on man-in-the-middle proxy~\cite{mitmproxy}. \tool intercepts outbound HTTPS traffic at an App's runtime, applies provider-specific fingerprinting rules to extract candidate credentials, and actively validates each credential against its provider's API.
Unlike prior work that focused on detecting hardcoded secrets in application binaries~\cite{LeakyAppSchmidt2025, ibrahim2025lmscout}, 
our study classifies each leaked credential by App category, leakage pattern, provider, and exploitability, and examines the anti-interception mechanisms and remediations deployed by developers. We investigate the following research questions:

\begin{itemize}
  \item \textbf{RQ1:} \textit{How prevalent is LLM API credential leakage in iOS Apps?} We analyze the landscape of API key leakage across iOS Apps, including the distribution across LLM service providers, App category, and user adoption levels (e.g., user rating counts).

  \item \textbf{RQ2:} \textit{How do API credentials leak in iOS Apps?} We examine the authentication and credential transmission mechanisms in intercepted network traffic to characterize the underlying leakage patterns, and further validate the exploitability of each leaked credential to assess its real-world security impact.

  \item \textbf{RQ3:} \textit{What protection mechanisms do developers adopt to prevent credential interception?} We analyze the commonly adopted anti-interception techniques in securing LLM API credential transmission, evaluating their effectiveness in preventing credential recovery from network traffic.

  \item \textbf{RQ4:} \textit{Do developers fix the credential leakage after responsible disclosure?} We re-evaluate the leaked credentials 90 days after disclosure to assess whether developers have taken appropriate remediation actions and to measure the effectiveness of the remediation efforts.
\end{itemize}

 Our empirical analysis reveals several findings:

\begin{enumerate}
  \item \textbf{Leakage is widespread.} \leakedTotal{} of \totalTestApps{} Apps (\testedExploitRate{}) expose LLM API credentials in network traffic, spanning at least ten LLM providers and five cloud platforms. Among these, \leakedFull{} (\leakedFullPct{}) are confirmed fully exploitable. The vulnerability affects Apps across \leakedCategories{} categories, with Health \& Fitness exhibiting the highest leakage rate (\healthLeakRate{}) and popular Apps with over \ratingMax{} ratings among those affected.

  \item \textbf{Three distinct leakage patterns emerge.} \jwtPct{} of the apps leaked JWT-based tokens that enabled LLM API service abuse, and \noAuthPct{} of the apps did not even require authentication to access backend services, rendering them directly exploitable. Meanwhile, \directKeyPct{} of the apps exposed plaintext API keys, with 47\% of affected apps simultaneously leaking proprietary system prompts. These findings indicate that developers still lack the domain knowledge to securely configure LLM API keys, and even those who adopt backend proxy architectures often fail to implement proper authentication and authorization mechanisms.

  \item \textbf{Client-side defenses are rarely deployed and often ineffective.} \mechPct{} of Apps implement any anti-interception mechanism. However, the most popular single defense (HTTP proxy bypass) still has a \proxyBypassRate{} bypass rate. In contrast, the multi-defense mechanism deployments exhibit a \multiBypassRate{} bypass rate, representing a \multiReduction{} $\times$ reduction, yet only 30\% of protected Apps adopt such combined strategies.

     \item \textbf{Remediation after disclosure remains critically low.} \retestRevokedPct{} of vulnerable Apps showed evidence of remediation after three months of responsible disclosure. Among the remaining \retestStillExploitable{} exploitable cases, \retestYesStill{} of Apps with \textit{Full} exploitability are unauthenticated backends with no credentials to revoke, while \retestLimitStill{} of Apps with \textit{Limited} exploitability remain vulnerable due to flawed authentication implementations, including missing JWT expiry claims, unenforced token lifetimes (up to 100 years), and servers that accept expired credentials.
\end{enumerate}





%% file: background.tex
\section{Background}
In this section, we introduce the authentication used by LLM APIs and the existing methodologies for analyzing iOS Apps.

\subsection{LLM API Authentication}

\noindent\textit{Static API Keys.} LLM providers (e.g., OpenAI, Google Gemini, Anthropic) authenticate requests using static API keys~\cite{openai_api_keys}. These long-lived bearer tokens (e.g., \texttt{sk-proj-...}) do not expire automatically and remain valid until revoked. During execution, Apps include API keys in HTTP request headers (Figure~\ref{fig:keyleakage}), making them directly observable and susceptible to interception. 

\noindent\textit{Short-Lived Tokens.} To reduce API key exposure, developers may build a backend proxy to mediate between the mobile app and the LLM provider; the proxy issues short-lived JSON Web Tokens (JWT) to the client instead of exposing the raw API key~\cite{rfc7519}. A JWT encodes claims (e.g., issuer, expiration, scope) and is validated by the proxy before forwarding requests to the LLM provider. However, if client authentication is not enforced, intercepted requests can be replayed to obtain unauthorized access.

Our study captures both directly exposed API keys and unauthenticated backend proxies by replaying intercepted requests. 

\subsection{Security Analysis for iOS Apps}

Apple's App Store distributes iOS apps as encrypted iOS Package Archive (IPA) bundles~\cite{apple_ipa_bundle} protected by FairPlay Digital Rights Management (DRM) ~\cite{fairplay}. FairPlay encrypts the App binary at distribution time and decrypts it on-device at load time using device-specific keys, preventing direct static inspection of the executable. As a result, many binary analysis techniques developed for Android, where APK contents are readily decompilable~\cite{weiAppSecretAndroid2025, zhouDevelopercredentialsAndroid2015, LeakyAppSchmidt2025}, cannot be directly applied to iOS Apps.
Therefore, to analyze iOS Apps, researchers have adopted two main methodologies:

\textbf{Static Binary Analysis via Device Jailbreaking.} iOS compiles Swift and Objective-C to native ARM64 machine code, so no decompiler can fully recover readable source code. Static analysis of iOS apps~\cite{PiOS_Egele2011, CRiOSOrikogbo2016, ihunterLiu2024USenix, LeakyAppSchmidt2025, iRiS_Deng2015, DominguezAlvareziOSLibKit2023fse} therefore requires a jailbroken device to decrypt and extract the FairPlay-protected binary from memory using tools such as \texttt{frida-ios-dump}~\cite{frida_docs} or \texttt{Clutch}~\cite{clutch_ios_dumper}. Once decrypted, researchers apply various techniques on the raw Mach-O binary. For instance, 
Feichtner et al.~\cite{ioscryptomisuseFeichtner2018} lifted ARM64 binaries to LLVM intermediate representation and applied static program slicing to detect cryptographic API misuse. More recently, Schmidt et al.~\cite{LeakyAppSchmidt2025} used regex pattern matching on decrypted binaries to detect hardcoded secrets across Android and iOS apps, and Liu et al.~\cite{ihunterLiu2024USenix} applied flow-sensitive taint analysis to detect privacy-violating data flows in third-party SDKs. 
However, these approaches require a jailbroken device (unavailable for all iOS versions) and cannot confirm whether extracted credentials are still active or exploitable.

\textbf{Dynamic Runtime Behavior Analysis.}
Due to the limitations of static analysis, prior work analyzes iOS Apps at runtime to detect information leakage~\cite{LalaineXiao23, RenPiiNetworktraffic2016, AndreaPrivacyDiffTest2017, MendozawebtoMobile2018}. 
They leverage runtime instrumentation frameworks such as Frida~\cite{frida_docs} to hook system APIs and trace sensitive data flows on live devices. For example, Xiao et al.~\cite{LalaineXiao23} used Frida to audit iOS system API invocations for privacy label compliance. However, this approach still requires a jailbroken device.
As a result, the majority of the work further employs network traffic interception~\cite{RenPiiNetworktraffic2016, AndreaPrivacyDiffTest2017, MendozawebtoMobile2018} to capture application-layer communications. This approach uses MITM proxies (e.g., \texttt{mitmproxy}~\cite{mitmproxy}) by installing a custom root certificate, avoiding jailbreaking and binary decryption.

Our study adopts network traffic analysis as the primary methodology because it does not require jailbreaking and directly observes credentials as they are transmitted, regardless of how they are stored, obfuscated, or constructed within the application binary.

%% file: methodlogy/overview.tex
\section{Methodology}\label{sec:methodology}

\begin{figure*}[hbtp]
    \vspace{-1.5em}
    \centering
\includegraphics[width=0.90\textwidth]{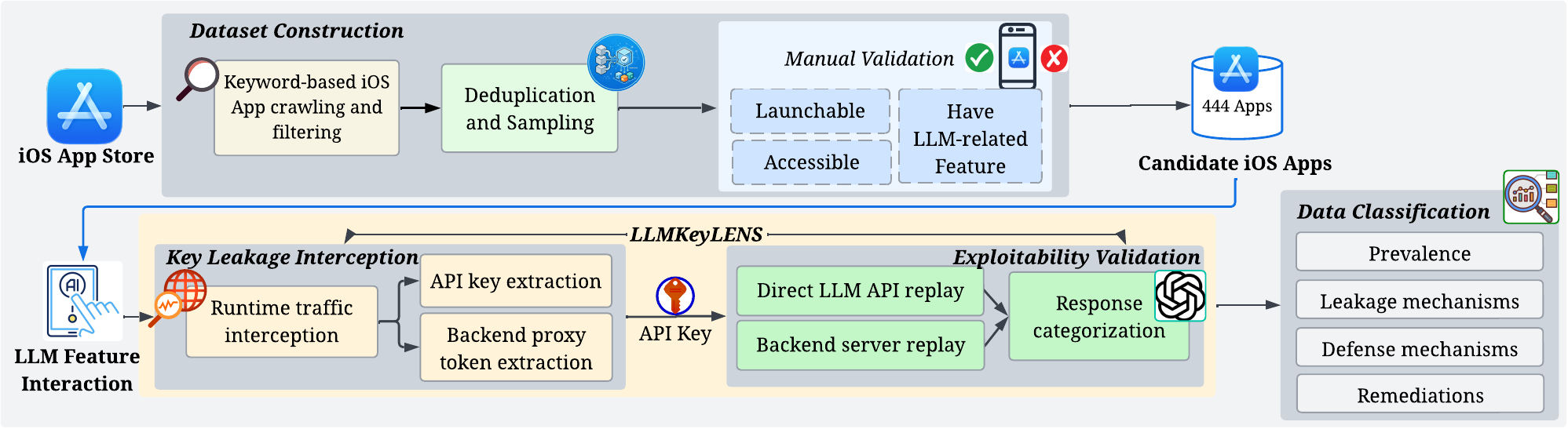}
    \vspace{-1em}
    \caption{Overview of Data Collection and Classification Workflow}
    \label{fig:overview}
    \vspace{-1em}
\end{figure*}

Our methodology includes four phases (as shown in Figure~\ref{fig:overview}). We first construct a curated dataset of LLM-integrated iOS applications through App Store crawling and manual screening (Section~\ref{sec:data-collection}). We then intercept runtime network traffic to extract candidate credentials (Section~\ref{sec:ui-testing}). Next, we validate credential exploitability and initiate responsible disclosure (Section~\ref{sec:validation}). Finally, we classify each application by leakage prevalence (\textbf{RQ1}), transmission mechanism (\textbf{RQ2}),  anti-interception defenses (\textbf{RQ3}), and remediation effectiveness  (\textbf{RQ4}).

\input{methodlogy/datasetCollection}

\input{methodlogy/analysisPipeline}

\input{methodlogy/credentialValidation}

\input{methodlogy/dataLabeling}

%% file: methodlogy/datasetCollection.tex
\subsection{Dataset Construction}
\label{sec:data-collection}
We focus on free iOS Apps that integrate LLM functionality and can be executed without pre-paid download, as our dynamic analysis requires actively exercising each App's LLM features at runtime.
We first designed search queries in three groups: (i) model names (e.g., GPT, ChatGPT, Claude, Gemini, LLaMA, Mistral, DeepSeek, Anthropic), (ii) AI functionality keywords (e.g., AI assistant, AI chatbot, AI writer), and (iii) combined queries pairing model names with functionality (e.g., "GPT + Coding Assistant", "Gemini+ Writer"). For each query, the API returns matching Apps along with their metadata (App name, description, developer, category, Track~ID, and Bundle~ID). The initial crawl yielded \totalCollectApp candidate Apps.
Next, we remove duplicated Apps using Track~ID (Apple's numeric App identifier) and Bundle~ID (the developer-assigned reverse-domain identifier). An App was considered a duplicate if either of its identifiers matched an existing entry; we obtained \totaldupApp{} unique Apps after de-duplication on October 25, 2025.

Since iOS Apps require interactive Apple ID authentication when installed through the App Store, which cannot be fully automated, we hired six undergraduate computer science students to conduct manual validation, each compensated at \$10 per hour for up to 30 hours. 
Due to the high cost of manual effort, we sampled 1,092 Apps from \totaldupApp{} using stratified random sampling~\cite{cochran1977sampling} to obtain a representative subset for evaluation, yielding a 99\% confidence level with a $\pm$3.5\% margin of error. 
To ensure labeling reliability, a shared test account was provided for apps requiring registration. Six evaluators were divided into three pairs, each cross-validating 364 apps on personal devices running iOS 18 against standardized inclusion criteria.  
\begin{enumerate}[leftmargin=*]
    \item \textit{Download failure} (45 apps excluded): The App could not be downloaded from the App Store at the time of evaluation (e.g., removed by the developer, restricted to other regions, or flagged by Apple).

    \item \textit{Installation or launch failure} (63 apps excluded): The App was successfully downloaded but could not be installed on the test device or crashed immediately upon launch without reaching a functional state.

    \item \textit{Accessibility} (318 apps excluded): The App could not be exercised due to one or more access barriers: (a)~mandatory paid subscription or in-app purchase required to unlock AI features, (b)~registration requiring invitation codes, enterprise credentials, or geographic restrictions, (c)~user-supplied API keys: Apps that require users to input their own LLM API keys.

    \item \textit{LLM functionality} (222 apps excluded):  After installation, the evaluator explored the App for up to five minutes following a standardized interaction protocol (navigating all visible screens, tapping interactive elements, and attempting text input). Apps without observable user interaction or content-generation features were excluded.
\end{enumerate}

For Apps where the two evaluators disagreed on the exclusion decision, a third evaluator (the first author) re-evaluated the App and resolved the disagreement through majority vote. Disagreements occurred in 46 of \totalevaApp{} Apps (4\%), predominantly on the boundary between criteria~(3) and~(4). After applying all exclusion criteria, the final dataset consists of \textbf{\totalTestApps{} LLM-integrated iOS Apps} with confirmed, exercisable LLM features.

%% file: methodlogy/analysisPipeline.tex
\subsection{API Key Leakage Interception}\label{sec:ui-testing}

Given the \totalTestApps{} Apps with confirmed LLM functionality from Section~\ref{sec:data-collection}, we designed \tool{} to intercept network traffic at runtime and detect API key leakage. The pipeline consists of (1)~traffic interception and capture, (2)~credential extraction through provider-specific pattern matching, and (3)~backend proxy detection for indirect leakage. Figure~\ref{fig:overview} illustrates the overall workflow.

\textbf{Traffic Interception.}
For each App, a tester installs the app on a physical iOS device configured to route all HTTP/HTTPS traffic through a mitmproxy~\cite{mitmproxy} instance on the host machine. A custom root CA certificate is installed in the iOS system trust store to enable TLS decryption. The tester then navigates to the App's LLM-powered feature and sends a predefined test prompt to trigger an API call. A mitmproxy addon script monitors all traffic and automatically records request--response pairs containing the test prompt, capturing the full URL, headers, and body as structured logs for credential extraction and validation (Section~\ref{sec:validation}). To parallelize testing across the \totalTestApps{} Apps, we run five mitmproxy instances on ports 8080--8084, each handling a partition of the App corpus.

However, many Apps bypass the iOS system HTTP proxy via low-level networking APIs (e.g., \texttt{NWConnection} from Apple's Network framework) that ignore the system proxy settings. Apps that produce no captured traffic or crash under the proxy configuration are flagged as \textit{proxy-bypassing}. We further retest these Apps via VPN-based transparent interception: all device traffic is routed through a VPN client (Potatso~\cite{potatso}) to the host machine, where mitmproxy operates in transparent proxy mode at the network layer, capturing traffic regardless of the App's proxy settings. 

In total, we captured \textbf{427/\totalTestApps{}} HTTP/HTTPS requests. 
We then extracted and analyzed embedded credentials as described.

\textbf{Credential Extraction from Direct API Calls.}\label{sec:credential-Extraction}
We extract candidate API keys by matching each request's endpoint domain, credential location (header, URL parameter, or body), and key prefix against provider-specific rules (e.g., \texttt{api.openai.com/v1/*} with an \texttt{sk-} prefixed key in the \texttt{Authorization} header for OpenAI, \texttt{generativelanguage.googleapis.com/v1/*} with an API key for Gemini and \texttt{api.anthropic.com/v1/*} with \texttt{x-api-key} for Claude). Applying these rules, we identified candidate API keys from \directApps{} Apps that made direct calls to known LLM provider endpoints.

\textbf{Backend Proxy Detection.}\label{sec:proxyTokenExtraction}
Not all credential leakage occurs through direct calls to LLM provider endpoints; some Apps route LLM requests through a developer-controlled backend proxy. To detect this pattern, we inspect the captured responses from all recorded HTTP traffic, including requests directed at non-provider domains. A response is classified as a proxied LLM response if it contains (i)~a \texttt{model} field matching a known LLM model identifier (e.g., \texttt{gpt-4}, \texttt{gemini-pro}), or (ii)~a \texttt{usage} object containing token-count fields (e.g., \texttt{prompt\_tokens}, \texttt{completion\_tokens}). When such a response is identified, the corresponding request is examined for authentication credentials, including Authorization headers, API key parameters, and custom authentication fields. If the request lacks any form of authentication, the endpoint is flagged as an \textit{unauthenticated LLM proxy}, indicating an exploitable inference endpoint that requires no credentials. Through this response-based detection, we identified an additional 355 Apps routing LLM requests through backend proxies. 

Our pipeline identified \textbf{421 candidate Apps} with potential API credential leakage, while 6 apps include no API key when calling the LLM provider directly.

%% file: methodlogy/credentialValidation.tex
\subsection{Exploitability Validation}\label{sec:validation}

To confirm that the extracted credentials are actually exploitable, we performed active validation testing for each extracted credential.


\textbf{Experiment Setup.} All experiments were conducted on a MacBook Pro with Apple M-series chip running macOS Sequoia, connected to a WiFi network shared with the test devices. For each candidate credential extracted during Phase 2, \tool automatically replays the captured request against the corresponding backend endpoint with a benign test payload, preserving the original request structure (URL, headers, and authentication fields) while substituting only the message content. To minimize unintended impact, each validation is constrained to a single request per credential with minimal resource consumption (e.g., setting max\_tokens=10 for LLM inference endpoints, or querying zero-cost metadata endpoints such as model listing APIs where available). No attempt was made to access user data, conversation histories, or account information beyond confirming whether the credential was accepted by the backend.

We classify each validated credential into three categories based on the backend response. A credential is classified as \textbf{\textit{actively exploitable}} if the backend returns a successful response with well-formed output, such as a valid inference result or a complete model listing, confirming that the key grants functional API access. A credential is classified as \textbf{\textit{restricted}} if the backend acknowledges the credential as valid but denies the request due to insufficient permissions (HTTP 403/401), exceeded quota (HTTP 429), billing suspension (HTTP 402), region-based access control, or if the application employs transport-layer protections that prevent credential replay (e.g., encrypted communication), indicating that the credential is authentic but not practically exploitable through our pipeline. A credential is classified as inactive if (i) the provider rejects it with an authentication failure response (HTTP 401 or 403) or an error indicating deactivation, or (ii) the credential cannot be validated through standard HTTP replay, such as when the original transmission occurs over a non-reproducible protocol (e.g., WebSocket).

For 421 apps with credential leakage, our validation confirmed that \textbf{\leakedTotal{}} apps remain \textit{actively exploitable}, returning valid LLM responses. The remaining 139 apps could not be exploited or actively validated: 66 had servers that were no longer operational, 44 were \textit{restricted}, and 29 were \textit{inactive}.

\noindent\textbf{Responsible Disclosure.}
We disclosed all \leakedTotal{} apps with exploitable credentials via their App Store contact channel, providing: the specific vulnerability description, sanitized traffic evidence, remediation recommendations (server-side proxy migration, key revocation), and a 90-day response timeline. 
As of the time of writing, \retestRevoked{} Apps have remediated the reported leakages, while \retestStillExploitable{} Apps remain exploitable with unchanged credentials.




%% file: methodlogy/dataLabeling.tex
\subsection{Data Classification}\label{sec:data-classification}

We classify the collected metadata corresponding to each App along multiple dimensions to answer our research questions. 

\noindent\textbf{RQ1 (Prevalence).} For each App with confirmed leakage, we linked the App to its App Store category and user adoption metrics from the metadata collected during Phase~1. 
Then, we classify each leaked credential by its LLM service provider and backend infrastructure using a combination of automated pattern matching and manual validation. Specifically, we match request endpoint domains, API key prefixes, and authentication token formats against predefined patterns to identify known LLM providers and infrastructure services. Based on this analysis, we categorize apps into three groups: (1) direct LLM API services, where requests are sent to known providers (e.g., OpenAI, Google Gemini); (2) cloud backends, where requests are routed through infrastructure services (e.g., Cloudflare Workers, Firebase, Google Cloud Run, Heroku); and (3) third-party services, including custom APIs and custom AI services that proxy or wrap LLM functionality. Apps that do not match any known patterns are labeled as \textit{unidentified}. This process enables us to analyze the distribution of leaked credentials across providers and backend infrastructures.

\noindent\textbf{RQ2 (Leakage Mechanisms).} We classify leakage mechanisms based on the integration patterns observed during key interception. Specifically, for each \textbf{\textit{actively exploitable}} credentials, we determine whether a request leaks credentials through: (i) Plaintext LLM API Key, when a provider credential (identified via provider-specific key prefixes) is included in requests sent directly to a known LLM endpoint; (ii) JWT/Bearer Token, when a bearer token is included in requests to a non-provider backend that proxies LLM access; and (iii) No Authentication, when no credential is present but LLM inference is successfully invoked via a backend proxy. We further classify the actively exploitable cases based on the observed leakage pattern. Plaintext LLM API Key and No Authentication are labeled as \textit{Full} exploitability due to unrestricted access, while JWT/Bearer Token is labeled as \textit{Limited} exploitability due to API constraints and token expiration.

\noindent\textbf{RQ3 (Defense Mechanisms).} 
\mechApps{} Apps yield no credential-bearing traffic during interception despite functioning normally on-device, indicating the presence of client-side defenses. We label the defense type based on the observable failure mode at the interception layer:
\begin{itemize}
\item \emph{HTTP proxy bypass}: Traffic does not appear in the MITM proxy, but is recovered by the VPN-layer capture.
\item \emph{Custom encryption}: Traffic is captured but the payload is encrypted beyond standard TLS, rendering credential extraction infeasible.
\item \emph{WebSocket}: The App communicates exclusively via WebSocket, bypassing HTTP-based interception.
\item \emph{Anti-detection}: The App refuses to launch or disables LLM features due to environment fingerprinting.
\item \emph{Network request failure}: The App detects the interception environment and silently stops issuing network requests.
\item \emph{Packet capture failure}: Neither the MITM proxy nor the VPN-layer capture produces any traffic.
\end{itemize}
Apps deploying more than one mechanism are classified as multi-mechanism deployments. Two authors independently labeled each protected App, with disagreements resolved through discussion.

\noindent\textbf{RQ4 (Remediation Outcomes and Insecure Coding Practices).} We re-evaluated the \textbf{\textit{actively exploitable}} leaked credentials 90 days after disclosure and assessed whether each credential remained active. We then classified the outcomes based on whether the credential was revoked, rotated, or remained unchanged, as well as whether access controls were strengthened or unchanged. This classification allows us to characterize remediation behaviors and identify recurring insecure coding practices.

%% file: experiment/expsetting.tex
\section{Major Findings}\label{sec:results}

In this section, we present our results and discuss the key findings corresponding to our research questions.

\input{experiment/result1}
\input{experiment/result2}

\input{experiment/result4}
\input{experiment/result3}



%% file: experiment/result1.tex
\subsection{Prevalence of LLM API Key Leakage}

Among \totalTestApps{} Apps, \noLeak{} Apps showed no exploitable leakage after evaluation, while \textbf{\leakedTotal{} Apps (\leakedTestPct{})} are validated with credential leakage. Specifically,  \leakedFull{} (\leakedFullPct{}) are fully exploitable, either through plaintext API keys that grant direct provider-level access or through unauthenticated backend proxies that require no credentials. The remaining \leakedLimited{} exhibit limited exploitability, where access is constrained to a backend proxy's API surface and subject to token expiration (Section~\ref{sec:rq2}). 


\subsubsection{Landscape of Leakage Among Different Apps' Categories}
\begin{figure}[htb]
    \centering
    \includegraphics[width=0.43\textwidth]{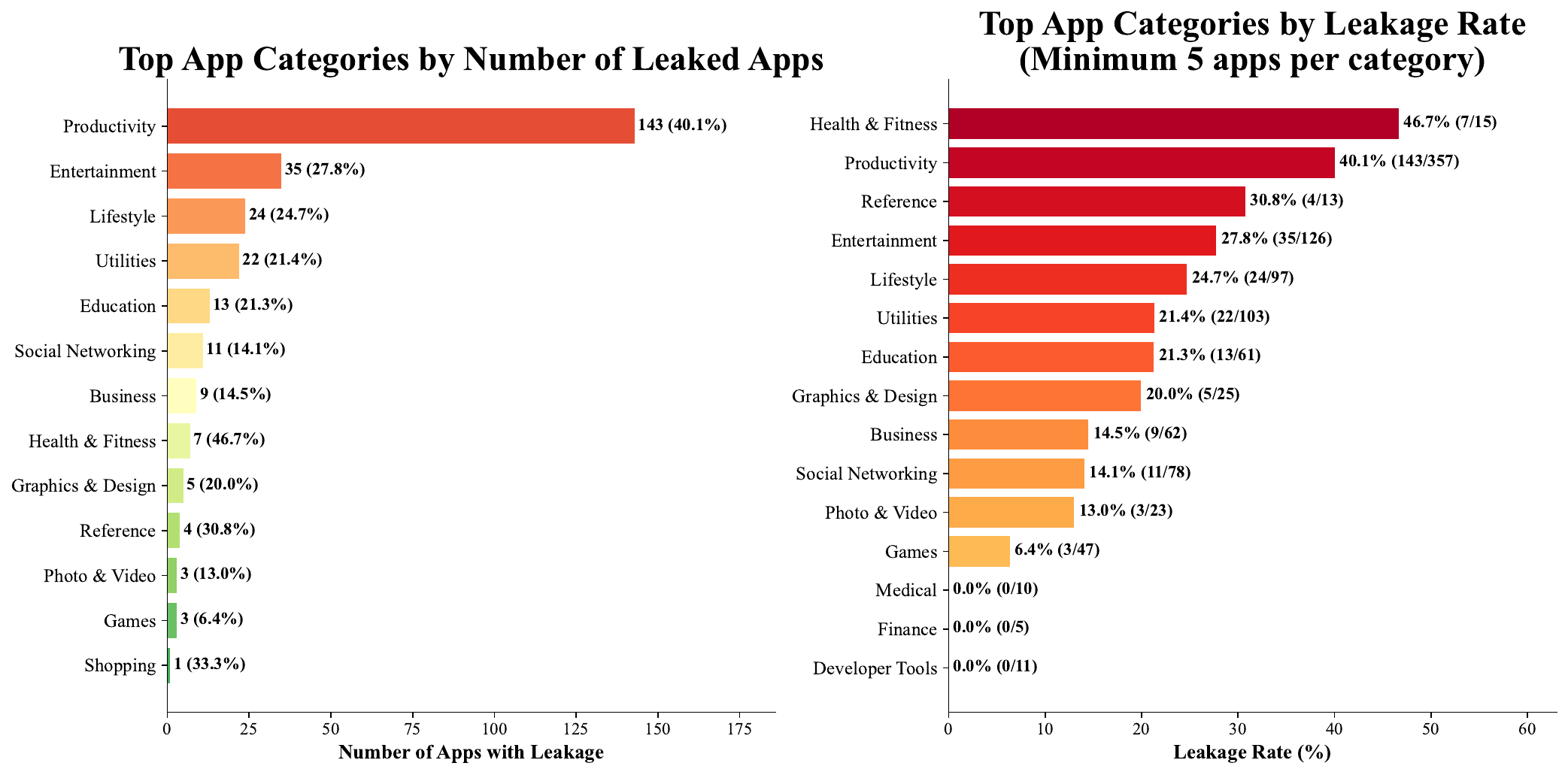}
    \caption{Distribution of LLM API key leakage across iOS app categories. The left panel shows absolute numbers; the right panel shows leakage rates for categories with $\geq$5 apps.}
    \vspace{-1em}
    \label{fig:Category}
\end{figure}

The distribution of vulnerable Apps spans \leakedCategories{} distinct categories as illustrated in Figure~\ref{fig:Category}. Productivity Apps account for the largest share with \prodLeaked{} vulnerable apps (\prodLeakedPct{} of all leaked Apps), followed by Entertainment (\entLeaked{} apps, \entLeakedPct{}) and Lifestyle (\lifeLeaked{} apps, \lifeLeakedPct{}). The leakage rates within each category are shown on the right side of Figure~\ref{fig:Category}, Health \& Fitness Apps exhibit the highest vulnerability rate at \healthLeakRate{} (\healthLeaked{} out of \healthTotal{} apps), followed by Productivity at \prodLeakRate{} (\prodLeaked{} out of \prodTotal{} apps) and Reference at \refLeakRate{} (\refLeaked{} out of \refTotal{} apps). Finance and Medical categories exhibit no leakage, despite containing \finTotal{} and \medTotal{} Apps, respectively.

The high leakage rate in Productivity apps likely stems from their heavy reliance on LLM-powered features (e.g., writing assistance, note-taking, translation), which often involve client-side API calls that expose credentials. In contrast, the absence of leakage in Finance and Medical apps may reflect stricter regulatory requirements and security practices, as well as less frequent use of LLMs or more secure integration designs.

\begin{figure}[htb]
    \centering
    \includegraphics[width=0.42\textwidth]{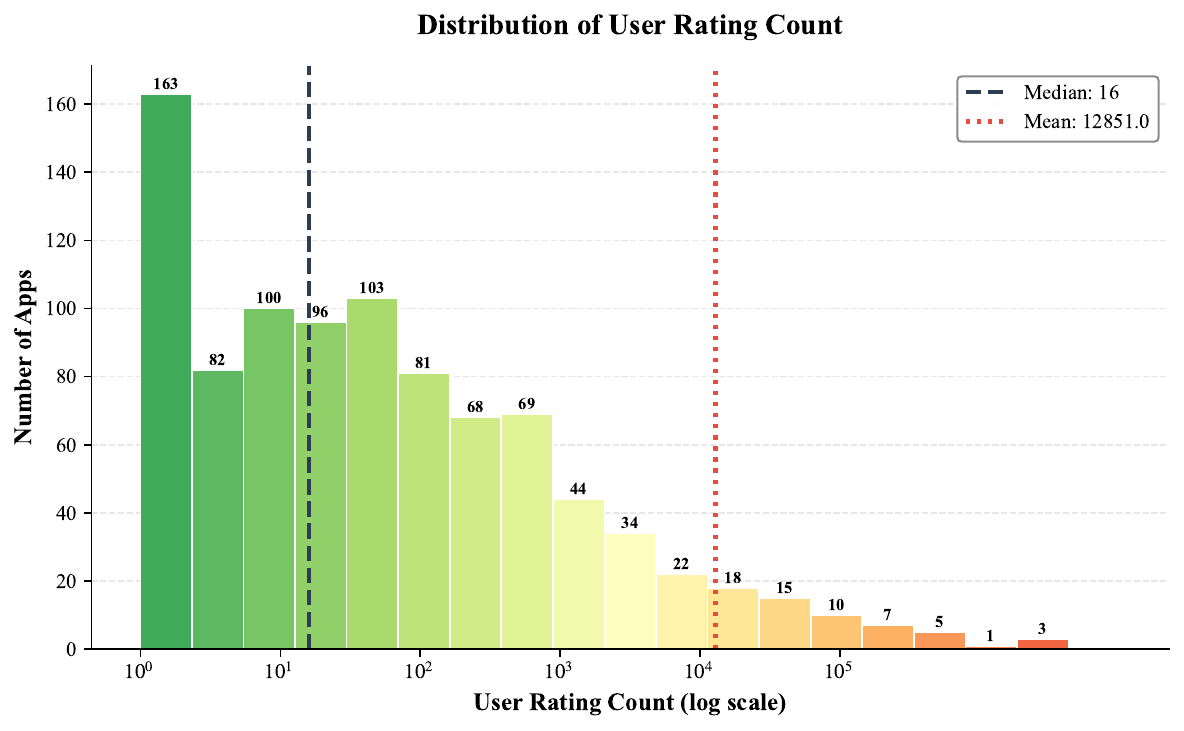}
    \caption{Distribution of user rating counts for vulnerable iOS Apps.}
    \label{fig:Count}
\end{figure}

 We further analyzed user adoption metrics for affected Apps. As shown in Figure~\ref{fig:Count}, the median rating count is \ratingMedian{}, indicating that many vulnerable Apps are niche products with limited user bases. However, the mean rating count of \ratingMean{} indicates that widely adopted Apps also suffer from credential-leakage vulnerabilities. Notably, \ratingOverKPct{} of vulnerable Apps have more than 1,000 user ratings, with the most popular receiving over \ratingMax{} ratings, suggesting that credential leakage affects Apps across all adoption levels.

\finding{LLM API key leakage is a widespread and systemic issue in the iOS ecosystem, affecting \leakedPct{} of analyzed Apps across diverse categories and developer types. The vulnerability's impact extends from niche Apps to popular apps with hundreds of thousands of users.}

\subsubsection{Landscape of Credential Leakage Across Different LLM Providers}
We categorized the \leakedTotal{} Apps with confirmed credential leakage into three groups based on their backend architecture (Table~\ref{tab:vendor-distribution}): (1) Apps that directly invoke LLM provider APIs (\directApps{} apps, \directPct{}), (2) Apps that route requests through cloud platform backends (\cloudApps{} apps, \cloudPct{}), and (3) Apps that communicate with custom developer-maintained backends (\customApps{} apps, \customPct{}).

\begin{table}[htb]
\centering
\footnotesize
\caption{Distribution of \leakedTotal{} leaked Apps by backend architecture and credential exposure type.}
\vspace{-1em}
\label{tab:vendor-distribution}
\begin{threeparttable}
\begin{tabular}{L{.09\textwidth}|L{.12\textwidth}|L{.052\textwidth}|L{.065\textwidth}|L{.06\textwidth}}
\toprule
\textbf{Category} & \textbf{Provider} 
& \multicolumn{3}{c}{\textbf{Leakage Patterns}} 
 \\\cline{3-5}
& & \textbf{Plaintext Key (Full)}  & \textbf{JWT/Bearer (Limited)} & \textbf{No Auth (Full)} 
\\
\toprule
\multirow{3}{*}{\makecell[l]{\textit{Direct LLM API} \\ (\directKeyApps{} Apps)}} &
 OpenAI (42)            & 42 & 0 & 0  \\
& Google Gemini (7)     & 7  & 0 & 0   \\
& Others (11)\tnote{a}   & 5  & 6 & 0  \\
\toprule
\multirow{4}{*}{\makecell[l]{\textit{ Cloud Platform} \\ (\cloudApps{} Apps)} }
& Firebase (27)          & 0  & 20 & 7   \\
& Google Cloud Run (11)  & 0  & 4  & 7    \\
& AWS (5)               & 0  & 0  & 5    \\
& Others (24)\tnote{b}   & 0  & 14 & 10  \\
\toprule
\textit{Custom Backend} (\customApps{} Apps)       & N/A  & 0  & 92 & 63  \\
\bottomrule
\textbf{Total (\textbf{\leakedTotal{}})}  &       & \textbf{\directKeyApps{}}  & \textbf{\jwtApps{}} & \textbf{\noAuthApps{}}  \\
\bottomrule
\end{tabular}
\begin{tablenotes}\scriptsize
\item[a] Baidu ERNIE (3), OpenRouter (2), DeepSeek (1), Mistral (1), Poe (1), Zhipu AI (1), DeepAI (1), Writesonic (1).
\item[b] Supabase (7), Azure (3), Heroku (2), Vercel (2), Railway (2), Fly.io (2), RapidAPI (2), Google App Engine (2), and 2 others.
\end{tablenotes}
\end{threeparttable}
\end{table}

We found \directKeyApps{} Apps that leak API keys through direct LLM API calls, where the App embeds credentials in requests sent directly to a provider's inference endpoint. Among these, OpenAI is the most frequently affected provider with \openaiApps{} Apps (\openaiPct{}), likely reflecting its dominant market share in the LLM ecosystem. Google Gemini follows with \geminiApps{} Apps, while the remaining providers (Baidu ERNIE, OpenRouter, DeepSeek, Mistral, among others) collectively account for the rest. These results indicate that many developers neglect secure client-side key management, and the concentration of leakage among widely-adopted providers disproportionately enlarges the attack surface for credential exposure.

Beyond direct LLM API calls, we identified \cloudApps{} (\cloudPct{}) Apps that route LLM requests through developer-controlled intermediaries hosted on cloud platforms. Specifically,  Firebase (\firebaseApps{} apps) and Google Cloud Run (\gcrApps{}) are the most common choices, followed by Supabase, AWS, and 14 other platforms. Theoretically, such backend proxy architectures are recommended by LLM providers as a best practice because they isolate API keys on the server side, preventing direct client-side exposure. However, our results reveal that credential leakage still occurs in these architectures through exposed backend authentication tokens, misconfigured access controls, or unauthenticated proxy endpoints, enabling LLM API abuse without ever obtaining the original provider key. 


Interestingly, the largest group (\customApps{} apps, \customPct{}) consists of Apps that use custom developer-maintained backends. These Apps invoke developer-owned domains (e.g., \texttt{api.example.com\\/v1/chat}) that proxy LLM requests through proprietary server infrastructure. Similar to the cloud platform backends, these custom backends also suffer from credential leakage through exposed authentication tokens or unauthenticated endpoints, which remain exploitable regardless of the underlying LLM provider.

\finding{Over half of leaked Apps (\customPct{}) route LLM traffic through custom developer backends, making provider-side mitigations alone insufficient. Cloud platforms and direct API services account for comparable shares of leakage (\cloudPct{} and \directPct{}, respectively), confirming that adopting a proxy architecture does not prevent credential exposure.}

%% file: experiment/result2.tex
\subsection{Leakage Mechanisms of LLM API Credentials in iOS Apps}\label{sec:rq2}

We categorize the \leakedTotal{} vulnerable applications by their leakage mechanism as shown in Table~\ref{tab:vendor-distribution}.
Among the \leakedTotal{} vulnerable Apps, \directKeyApps{} (\directKeyPct{}) transmit \textit{\textbf{plaintext LLM API keys}} directly to provider endpoints. In these cases, the API key is embedded in plaintext within the \texttt{Authorization} header or as a URL query parameter (e.g., Google Gemini), allowing us to recover the API key from a single intercepted request. This pattern represents the most severe form of credential leakage, as it requires no additional exploitation steps beyond passive network traffic interception. As illustrated in Listing~\ref{lst:real-directkey}, the request header contains both the OpenAI project key (\texttt{sk-proj-*}) and the organization identifier without any form of obfuscation or encryption. A single captured request provides an attacker with sufficient information to invoke any model under the developer's account with arbitrary prompts.
More critically, we observed that 28 out of \directKeyApps{} Apps also disclose system prompts within the \texttt{messages} array of the same request body. This co-exposure stems from the design of LLM provider APIs, which require both the authentication credential and the \texttt{messages} payload (e.g., system prompt) as parameters within a single API call. Therefore, when developers directly integrate with LLM provider APIs without a backend proxy, all of this information is inevitably embedded in client-side traffic. This indicates that an adversary can perform a compound attack from a single interception, gaining both unrestricted inference access and direct extraction of proprietary system prompts that encode the App's core business logic.

\begin{lstlisting}[
    language={},
    basicstyle=\tiny\ttfamily,
    frame=single,
    breaklines=true,
    columns=fullflexible,
    escapeinside={(@}{@)},
    float=h,
    caption={Plaintext OpenAI API key in request headers.},
    label={lst:real-directkey}
]
POST /v1/chat/completions HTTP/1.1
Host: api.openai.com
Headers:
  Content-Type:        application/json
  (@\hlt{Authorization:       Bearer sk-proj-****...****}@)
  (@\hlt{OpenAI-Organization: org-****...****}@)
  User-Agent:          [AppName]/2 CFNetwork/3860.100.1
Body:
  {"model": "gpt-5",
   "messages": [{"role": "user", "content": "..."}],
   "stream": true}
\end{lstlisting}

Surprisingly, we found that \noAuthApps{} Apps (\noAuthPct{}) route LLM requests through backend proxies \textbf{\textit{without authentication}}. In these cases, although developers correctly relocate the LLM API key to the server side, separating it from client-side logic, they fail to enforce access control on the proxy endpoint itself. As illustrated in Listing~\ref{lst:real-noauth}, the intercepted request contains no \texttt{Authorization} header, API key, or bearer token; nevertheless, the backend (implemented as a Google Cloud Function) processes the unauthenticated POST request and returns an LLM-generated response. This finding suggests that many developers lack sufficient understanding of server-side authentication practices. Such unauthenticated endpoints effectively create an open relay that we categorize as \emph{Fully Exploitable}, since an adversary can simply obtain the endpoint URL and request structure from an intercepted request and gain unrestricted access to the underlying LLM service without requiring any credentials.

\begin{lstlisting}[
    language={},
    basicstyle=\tiny\ttfamily,
    frame=single,
    breaklines=true,
    columns=fullflexible,
    escapeinside={(@}{@)},
    float=h,
    caption={Unauthenticated Cloud Function endpoint accepts arbitrary LLM requests.},
    label={lst:real-noauth}
]
POST https://us-central1-[project-id].cloudfunctions.net/chatStream
Headers:
  content-type: application/json
  user-agent: [AppName]/1 CFNetwork/3826.600.41
  (@\hlt{(no Authorization header)}@)
Body:
  {"data":{"model":"gpt-5.1","messages":[{"content":[{"type":"text","text":"Tell me a joke"}],"role":"user"}]}}
\end{lstlisting}

We observed \jwtApps{} Apps where leaked JSON Web Token (JWT) bearer tokens remain actively exploitable. Unlike direct LLM API key exposure or unauthenticated endpoints, these Apps employ backend proxy architectures where JWT tokens are issued by authentication services such as Firebase Authentication, custom OAuth providers, or proprietary backend systems. As shown in Listing~\ref{lst:real-jwt}, the JWT token in the \texttt{Authorization} header authenticates the client to a backend proxy that forwards the request to an LLM provider. While this design avoids direct exposure of the LLM API key, it requires developers to correctly configure authentication parameters such as token expiration time. In our evaluation, we found that many captured JWT tokens remained valid at the time of testing, allowing us to replay them against the same backend endpoint with arbitrary prompts. We classify this pattern as \emph{Limited Exploitability}, as access is constrained to the backend's API surface and tokens are eventually subject to expiration.


\begin{lstlisting}[
    language={},
    basicstyle=\tiny\ttfamily,
    frame=single,
    breaklines=true,
    columns=fullflexible,
    escapeinside={(@}{@)},
    float=t,
    caption={JWT bearer token authenticates to a backend proxy.},
    label={lst:real-jwt}
]
POST https://api.exh.ai/chatbot/v4/[app]/response
Headers:
  content-type: application/json
  x-app-id: com.[redacted]
  (@\hlt{authorization: Bearer eyJhbGciOiJIUzUxMiJ9.eyJ1c2Vy...6fY5af...}@)
Body:
  {"message": "Hello", "bot_id": "..."}
\end{lstlisting}

\finding{JWT-based leakage is the most prevalent category (\jwtPct{}), indicating that many developers adopt backend proxy architectures while still relying on client-side tokens that are easily interceptable. In contrast, plaintext API key exposure, though less frequent (\directKeyPct{}), incurs the highest risk due to unrestricted access. No-auth leakage has the lowest barrier to exploitation but is limited to the exposed endpoint's functionality.}

%% file: experiment/result4.tex
\subsection{Defense Mechanisms and Their Effectiveness Against Credential Leakage}\label{sec:rq3}
Among the \analyzableApps{} Apps for which traffic interception was attempted, \mechApps{} (\mechPct{}) implemented at least one mechanism to resist HTTP proxy interception. Table~\ref{tab:bypass} summarizes the distribution of anti-interception mechanisms observed across these \mechApps{} Apps. 
 Still, our analysis pipeline successfully bypassed \totalBypassed{} Apps. 
 


\textbf{Single Defense Mechanisms and Their Effectiveness.} 
HTTP proxy bypass is the most widely adopted single mechanism, deployed by \proxyBypassApps{} Apps. Despite its popularity, it proved to be the weakest defense: \proxyBypassBypassed{} of \proxyBypassApps{} Apps (\proxyBypassRate{}) were bypassed through \tool's VPN-based fallback, which captures traffic at the network layer regardless of proxy configuration. The remaining single mechanisms demonstrated substantially stronger resistance. \netFailApps{} Apps silently stopped issuing requests upon detecting the interception environment (\netFailBypassed{} bypassed). \packetFailApps{}, \encryptApps{}, \antiDetectApps{}, and \wsApps{} Apps employing packet capture resistance, custom encryption, anti-detection, and WebSocket (WS) channels, respectively, all achieved a 0\% bypass rate. These results indicate that HTTP proxy bypass is the only single mechanism with a non-trivial bypass rate.

\textbf{Combined Defense Mechanisms.}
\multiApps{} Apps deployed multiple anti-interception mechanisms simultaneously as listed in Table~\ref{tab:bypass}.
Compared with a single defense mechanism, combined defense mechanisms have a much lower overall bypass rate (\multiBypassRate{}  vs.  \singleBypassRate{}). In detail,   
HTTP proxy bypass serves as the base layer in most combinations, appearing in 40 of \multiApps{} multi-mechanism deployments (\proxyInMultiPct{}), but is consistently paired with complementary mechanisms that address its standalone weakness.  
The most common combination is HTTP proxy bypass with custom encryption (\proxyEncApps{} Apps), achieving a \proxyEncRate{} bypass rate, as the encrypted payload remained protected even when the proxy evasion was circumvented.
\wsProxyApps{} Apps adopted WebSocket with HTTP proxy bypass, with only \wsProxyBypassed{} leak credentials during our evaluation. This protection requires defeating both the non-HTTP protocol barrier and the proxy evasion simultaneously.
\wsProxyEncApps{} Apps deployed a three-layer defense combining WebSocket channels, HTTP proxy bypass, and custom encryption; and they are entirely resistant to our analysis pipeline. 

Our results indicate that HTTP proxy bypass is the most widely adopted anti-interception mechanism, yet also the least effective with a \proxyBypassRate{} bypass rate. In contrast, layered defenses combining proxy bypass with custom encryption or WebSocket channels reduce bypass rates to near zero. However, only around 10\% of evaluated Apps adopt such combined strategies, suggesting that developers overwhelmingly favor ease of implementation over robust credential protection. We recommend that LLM providers and platform documentation explicitly advocate defense-in-depth strategies rather than single-mechanism approaches.

\finding{Only \mechPct{} of evaluated Apps deploy anti-interception mechanism, indicating a significant gap between available defenses and actual developer practice. Multi-mechanism deployments achieve a \multiBypassRate{} bypass rate versus \singleBypassRate{} for single-mechanism approaches, a \multiReduction{}x reduction. HTTP proxy bypass, while appearing in \proxyInMultiPct{} of combined defenses, is effective only when paired with custom encryption or WebSocket channels. }
\vspace{-.5em}

\begin{table}[h]
  \centering
  \footnotesize
  \caption{Anti-interception mechanisms in \mechApps{} Apps.}
  \vspace{-1.5em}
  \label{tab:bypass}
  \begin{threeparttable}
  \begin{tabular}{lrrr}
    \toprule
    \textbf{Mechanism} & \textbf{Total} & \textbf{Bypassed} & \textbf{Rate} \\
    \midrule
    \multicolumn{4}{l}{\textit{Single mechanism (\singleApps{} apps)}} \\
    \quad HTTP proxy bypass          & \proxyBypassApps{} & \proxyBypassBypassed{} & \proxyBypassRate{} \\
    \quad Network request failure    & \netFailApps{} & \netFailBypassed{} & \netFailRate{} \\
    \quad Packet capture failure     & \packetFailApps{} &  0 &  0\% \\
    \quad Custom encryption          & \encryptApps{} &  \encryptBypassed{} &  \encryptRate{} \\
    \quad Anti-detection             & \antiDetectApps{} &  0 &  0\% \\
    \quad WebSocket (WS)       & \wsApps{} &  0 &  0\% \\
    \midrule
    \multicolumn{4}{l}{\textit{Multiple combined (\multiApps{} apps)}} \\
    \quad Proxy bypass + encryption  & \proxyEncApps{} & \proxyEncBypassed{} & \proxyEncRate{} \\
    \quad WS + proxy bypass   & \wsProxyApps{} & \wsProxyBypassed{} & \wsProxyRate{} \\
    \quad Proxy bypass + net.\ fail  & \proxyNetFailApps{} & \proxyNetFailBypassed{} & \proxyNetFailRate{} \\
    \quad WS + proxy + encryption    & \wsProxyEncApps{} &  0 &  0\% \\
    \quad WS + encryption     & \wsEncApps{} &  0 &  0\% \\
    \quad Other\tnote{$\dagger$}     & \multiOtherApps{} & \multiOtherBypassed{} & \multiOtherRate{} \\
    \midrule
    \textbf{Total}       & \textbf{\mechApps{}} & \textbf{\totalBypassed{}} & \textbf{\totalBypassRate{}} \\
    \bottomrule
  \end{tabular}
  \begin{tablenotes}\scriptsize
  \item[$\dagger$] WS + proxy bypass + net.\ failure (1),
    app failure + proxy bypass (1), proxy bypass + anti-detection (1),
    packet failure + anti-detection (1).
  \end{tablenotes}
  \end{threeparttable}
  \vspace{-2em}
\end{table}

%% file: experiment/result3.tex
\subsection{Remediation Outcomes and Persistent Insecure Coding Practices}

To assess whether developers conducted effective remediation after our responsible disclosure, we replayed the original intercepted requests 90 days after the disclosure process described in Section~\ref{sec:validation}. Table~\ref{tab:retest} summarizes the evaluation outcomes. 

Out of the \retestTotal{} applications retested, \retestRevoked{} (\retestRevokedPct{}) now return HTTP 401 or 403, indicating successful credential revocation or newly enforced access controls. Among the \retestYesRevoked{} remediated Full-exploitability applications, 15 revoked their plaintext LLM API keys (12~OpenAI, 3~Gemini) and 12 added authentication to previously unauthenticated proxy endpoints. For Limited-exploitability applications, \retestLimitRevoked{} now correctly reject expired JWT tokens, suggesting that token lifecycle management is effective when properly configured.
We also found that \retestUnreachable{} applications (\retestUnreachablePct{}) have become unreachable due to endpoint decommissioning or App Store delisting, indicating that some developers chose to eliminate the attack surface entirely rather than apply a targeted fix.


\begin{table}[hbtp]
\centering
\footnotesize
\caption{Remediation outcomes after responsible disclosure.}
\vspace{-1em}
\label{tab:retest}
\begin{tabular}{lrrrr}
\toprule
\textbf{Outcome} & \multicolumn{2}{c}{\textbf{Full}} & \multicolumn{2}{c}{\textbf{Limited}} \\
\cmidrule(lr){2-3}\cmidrule(lr){4-5}
 & \textbf{n} & \textbf{(\%)} & \textbf{n} & \textbf{(\%)} \\
\midrule
Still exploitable (HTTP 200)     & \retestYesStill{}  & 25 & \retestLimitStill{}  & 22 \\
Remediated (HTTP 401/403)        & \retestYesRevoked{} & 18 & \retestLimitRevoked{} & 38 \\
Rate limited (HTTP 429)          & \retestRateLimited{} & 6 & 0 & 0 \\
Endpoint unreachable             & 47 & 32 & 36 & 26 \\
Other server error               & 27 & 18 & 19 & 14 \\
\midrule
\textbf{Total}                   & \textbf{\retestFull{}} & \textbf{100} & \textbf{\retestLimited{}} & \textbf{100} \\
\bottomrule
\end{tabular}
\end{table}

However, we found \textbf{\retestStillExploitable{} applications (\retestStillExploitablePct{})} remain exploitable, indicating that the underlying insecure practices persist despite notification. These cases fall into two categories: 

\textit{\textbf{No Remediation (\retestYesStill{} Apps).}} \retestYesNoAuth{} of \retestYesStill{} applications are unauthenticated backend proxies, where disclosure alone did not prompt the structural change required to add server-side access control. The remaining \retestYesDirectKey{} include applications with unrotated LLM API keys. These results suggest that responsible disclosure is insufficient to drive remediation when the root cause requires architectural changes rather than credential replacement.

\textit{\textbf{Incorrect Implementation of Authentication (\retestLimitStill{} Apps).}} These applications were initially classified as Limited Exploitability on the premise that their JWT or bearer tokens would eventually expire. However, after 90 days, the captured tokens remained valid and exploitable. To understand why, we decoded all \retestLimitStill{} tokens and identified the root causes (Table~\ref{tab:retest-rootcause}), which fall into two categories. The first category comprises credentials that \emph{cannot} expire by design, including applications that use static bearer tokens with no expiry mechanism, issue JWTs without an \texttt{exp} claim, or require no authentication at all. The second category comprises credentials where an expiry policy \emph{exists} but is not enforced: some applications accept JWTs beyond their \texttt{exp} deadline, while others issue JWTs with excessively long lifetimes ranging from 100 to 365 days (one token is valid for 100 years). This finding stands in stark contrast to the \retestLimitRevoked{} applications that successfully remediated, all of which employed JWT-based authentication where the server correctly rejected expired tokens. The contrast demonstrates that the gap lies not in \emph{adopting} authentication mechanisms but in \emph{correctly implementing} them: \retestLimitStill{} of \retestLimited{} applications (\retestLimitStillPct{}) originally classified as Limited Exploitability are, in practice, persistently vulnerable as those with directly exposed API keys.


\finding{After responsible disclosure, \retestRevokedPct{} of vulnerable applications successfully remediated through credential revocation or access control enforcement. However, \retestStillExploitablePct{} remain exploitable due to either absence of remediation action (\retestYesStill{} apps) or fundamentally flawed authentication implementations (\retestLimitStill{} apps).}

\subsubsection{Case Study} To gain deeper insight into the leakage mechanisms and developer remediation practices, we conduct case studies on representative Apps from each leakage category.

\noindent\textbf{Case Study A: Unauthenticated Cloud Function.}
We illustrate this pattern with \textit{App~A}, a niche chat application whose backend is a Google Cloud Function. As shown in Listing~\ref{lst:case-noauth}, the request to \texttt{/openAIProxy} contains no \texttt{Authorization} header, API key, or bearer token, yet the body specifies the OpenAI model (\texttt{gpt-3.5-turbo}), a system prompt, and a token budget of 400. The server returns HTTP~200 with a full \texttt{chat.completion} response, consuming 168 tokens billed to the developer's OpenAI account. The developer correctly relocated the API key into the Cloud Function, but the proxy endpoint accepts requests from any client without access control. Ninety days after our disclosure, the endpoint remains fully open with identical behavior.

\begin{lstlisting}[
    language={},
    basicstyle=\tiny\ttfamily,
    frame=single,
    breaklines=true,
    columns=fullflexible,
    escapeinside={(@}{@)},
    float=htb,
    caption={Unauthenticated Cloud Function proxy.},
    label={lst:case-noauth}
]
POST /openAIProxy HTTP/1.1
Host: us-central1-[project-id].cloudfunctions.net
Headers:
  Content-Type: application/json
  (@\hlt{(no Authorization header)}@)
Body:
  {"model":"gpt-3.5-turbo",
   "messages":[{"role":"system","content":"You are ChatGPT, a large language model trained by OpenAI"}, {"role":"user","content":"Hello"}], "max_tokens":400} 
Response: HTTP 200
  {"model":"gpt-3.5-turbo-0125",
   "usage":{"prompt_tokens":21,"completion_tokens":168,"total_tokens":189}}
\end{lstlisting}

\begin{table}[htb]
\vspace{-1.5em}
\centering
\footnotesize
\caption{Root causes of persistent exploitability among the \retestLimitStill{} Apps.}
\vspace{-1.5em}
\label{tab:retest-rootcause}
\begin{tabular}{p{4.0cm}rr}
\toprule
\textbf{Root Cause} & \textbf{Apps} & \textbf{\%} \\
\midrule
Static/opaque token (no expiry) & \retestLimitStatic{} & 43 \\
Excessive JWT lifetime ($\geq$100\,d) & \retestLimitExcessive{} & 20 \\
Expired JWT accepted by server  & \retestLimitExpiredAccepted{} & 17 \\
JWT issued without \texttt{exp} claim & \retestLimitNoExp{} & 17 \\
Short-lived JWT (user session)  & \retestLimitSessionToken{} & 3 \\
\midrule
\textbf{Total} & \textbf{\retestLimitStill{}} & \textbf{100} \\
\bottomrule
\end{tabular}
\end{table}

\noindent\textbf{Case Study B: Flawed JWT Authentication.}
 We present two representative JWT failure modes that remained exploitable despite employing token-based authentication.

\noindent\textit{\underline{Excessive token lifetime.}} \textit{App~B}, a popular application with over 100K user ratings, uses a custom backend at \texttt{api.[redacted]/graphql} with HS256-signed JWTs. As shown in Listing~\ref{lst:case-jwt-tolan}, the token's \texttt{exp} claim is set to the year 2125, yielding a lifetime of 36,525~days (100.1~years). Although the developer adopted JWT authentication with proper signing, setting the expiry a century into the future renders token-based access control meaningless. The token remains valid 90~days after disclosure.

\begin{lstlisting}[
    language={},
    basicstyle=\tiny\ttfamily,
    frame=single,
    breaklines=true,
    columns=fullflexible,
    escapeinside={(@}{@)},
    float=htb,
    caption={JWT with 100-year lifetime.},
    label={lst:case-jwt-tolan}
]
JWT Payload (decoded):
{
  (@\hlt{"iat": 1763693717}@),   // 2025-11-20
  (@\hlt{"exp": 4919453717}@),   // 2125-11-21 (100.1 years)
  "aud": "USER_AUTH",
  "sub": "usr_[redacted]"
}
Token lifetime: 36,525 days
Response: HTTP 200
  {"data":{"chatMessageCreate":{"__typename":"ChatMessageOutput"}}}
\end{lstlisting}

\noindent\textit{\underline{Expired token accepted by server.}} \textit{App~C} uses Firebase anonymous authentication to issue short-lived JWTs (1-hour lifetime) and routes requests through a Zuplo API gateway to OpenAI. Listing~\ref{lst:case-jwt-janora} shows the decoded payload: the token was issued on 2025-11-14 at 13:06 with a 1-hour expiry--a theoretically sound design with appropriately short token lifetimes. However, when we replayed this token 128 days after expiration, the server returned HTTP 200 with a valid LLM response. Furthermore, the response headers revealed the upstream OpenAI organization identifier and project ID, which indicates that the server-side proxy does not perform proper token validation and inadvertently exposes exploitable credentials.

Our observations show that developers try to route LLM API calls through a server-side. However, they may still lack the security expertise to implement proxy authentication correctly. To protect LLM API key credentials, responsibility is shared across three stakeholders. Developers should route all LLM API calls through authenticated server-side proxies, ensure backend endpoints enforce access control, and revoke leaked keys immediately upon discovery. LLM providers should (i) explicitly document client-side integration as an insecure pattern, (ii) publish reference implementations of authenticated proxy architectures to lower the barrier for secure adoption, and (iii) introduce anomaly detection that flags API keys accessed from a high volume of distinct clients, proactively notifying affected developers upon detection. The iOS platform should integrate dynamic analysis tools such as \tool into the App Store review process to detect transmission-layer credential exposure before deployment.

\begin{lstlisting}[
    language={},
    basicstyle=\tiny\ttfamily,
    frame=single,
    breaklines=true,
    columns=fullflexible,
    escapeinside={(@}{@)},
    float=htb,
    caption={1-hour JWT accepted 128 days past expiry.},
    label={lst:case-jwt-janora}
]
JWT Payload (decoded):
{
  "iss": "securetoken.google.com/[project-id]",
  "aud": "[project-id]",
  (@\hlt{"iat": 1731585960}@),   // 2025-11-14 13:06 UTC
  (@\hlt{"exp": 1731589560}@),   // 2025-11-14 14:06 UTC (1-hour lifetime)
  "auth_time": 1731585960,
  "provider_id": "anonymous"
}
Token replayed: 2026-03-22 (128 days past expiry)
Response: HTTP 200
Response Headers (excerpt):
  (@\hlt{openai-organization: kaan2}@)
  (@\hlt{openai-project: proj\_mbZvm8...}@)
\end{lstlisting}
\vspace{-0.5em}

\finding{Mitigation remains ineffective primarily due to (1) unauthenticated backend proxies that forward credentials without verifying client identity, and (2) incorrect implementation of token-based authentication and validation.}

%% file: Related.tex
\section{Related Work}\label{sec:related-work}

\subsection{Empirical Studies on Credential Leakage}

Credential leakage has been extensively measured across multiple software ecosystems~\cite{githubkeyMeliMR19,dotfile2023,Soufiancloud2025,icseBasakNRW23Secrect, zhang2023don,LeakyAppSchmidt2025, ibrahim2025lmscout, liu2023prompt, LiuAndroidRuntime2018}. Meli et al.~\cite{githubkeyMeliMR19} found thousands of secrets leaked daily on GitHub with the majority remaining accessible after disclosure, while Krause et al.~\cite{Alexandergithub2023} studied developer strategies for handling secrets, finding that most leaks result from accidental commits and that detection tools such as secret scanners remain underutilized. 
Basak et al.~\cite{icseBasakNRW23Secrect} investigated the challenges developers face with checked-in secrets in software artifacts, identifying key barriers to secret management adoption. 
Zhou et al.~\cite{zhouDevelopercredentialsAndroid2015} were the first to study developer credential harvesting on Android, identifying over 1,000 apps that embedded plaintext cloud service credentials. Subsequent work broadened the scope to cloud-backed mobile apps~\cite{zuoCloudApp2019}, web-to-mobile API inconsistencies~\cite{MendozawebtoMobile2018}, and platform-specific credential leakage in WeChat mini-programs~\cite{zhang2023don}. Wei et al.~\cite{weiAppSecretAndroid2025} empirically studied how far app secrets are from being stolen on Android, analyzing the full attack chain from secret embedding to exploitation. 


Existing work also studies the credential leakage via LLM or LLM-integrated applications~\cite{zhou2025LLMJailbreak, brucato2024llmjacking, promptleakge24CCS, GulayY25LLM, chen2025understanding}. Specifically, Hou et al.~\cite{hou2025security} provided the first systematic security analysis of LLM app stores, examining over 6,000 applications and identifying prevalent vulnerabilities, including prompt injection and insufficient access control. Deng et al.~\cite{masterDengLLWZLW0L24} formalized prompt injection attacks against off-the-shelf LLMs, demonstrating that compromised access can be compounded with adversarial prompts to hijack application behavior and cause sensitive information leakages.

Our work differs from existing studies by targeting a new credential type (LLM API keys) in iOS apps, and further characterizes leakage mechanisms and exploitability rather than solely detecting the presence of secrets.


\subsection{Detection of Credential and Data Leakage}

A variety of techniques~\cite{FengPwdGithub, aseYangBLK22Notebooks, MengWXBW23WemintASE,asemincheckerWang0ZWJLCLZH024, Li25FirmwareLeakage, RenPiiNetworktraffic2016,nanPrivacyMobi2018, WangLPLL25detectionicse} have been proposed to detect credential and sensitive data leakage across different platforms. 
Meng et al.~\cite{MengWXBW23WemintASE} proposed Wemint, a taint analysis approach for detecting sensitive data leaks in WeChat mini-programs, and Wang et al.~\cite{asemincheckerWang0ZWJLCLZH024} developed MiniChecker to detect abusive permission request behavior that leads to data privacy risks. Ren et al.~\cite{RenPiiNetworktraffic2016} built ReCon for revealing PII leaks in mobile network traffic, and subsequent work improved detection through obfuscation-resilient differential analysis~\cite{AndreaPrivacyDiffTest2017}, semantics-driven learning-based discovery~\cite{nanPrivacyMobi2018}, and longitudinal tracking of PII leaks across app versions~\cite{renPiiinAndroid2018}. Beyond mobile apps, Li et al.~\cite{Li25FirmwareLeakage} recently addressed firmware leakage in IoT update processes by analyzing companion applications. At the source code level, Feng et al.~\cite{FengPwdGithub} applied deep learning to detect password leakage at the code semantics level across 4.6 million GitHub files, and Yang et al.~\cite{aseYangBLK22Notebooks} proposed static detection techniques for data leakage in computational notebooks, addressing a growing but underexplored attack surface.

Our empirical findings and \tool could serve as a foundation for tool builders to extend dynamic analysis pipelines with provider-specific credential identification and validity confirmation.


\subsection{Mitigation of Credential Leakage}


Despite these risks, mitigation efforts remain inadequate. Krause et al.~\cite{Alexandergithub2023} found that developers are aware of secret scanning tools but rarely adopt them, with most leaks attributed to accidental commits rather than ignorance. Yadmani et al.~\cite{Soufiancloud2025} reported that responsible disclosure of leaked cloud credentials yields low remediation rates, with many secrets remaining accessible months after notification. On the defensive side, certificate pinning has been studied as a mitigation against network interception; Pradeep et al.~\cite{pradeepCertPinning2022} conducted a comparative analysis of certificate pinning adoption in Android and iOS, finding that only a small fraction of apps implement it correctly, leaving most traffic susceptible to MITM interception.

Our study first measures developers' remediation responses following the responsible disclosure of the LLM API key leakage. We find that insecure coding practices, such as missing authentication on backend proxies and broken JWT implementations, persist even after notification.

%% file: Threats.tex
\section{Threats to Validity}

\textbf{Internal Validity.}
\tool relies on MITM proxy interception, so it can miss the Apps that employ HTTP proxy bypass and route traffic outside the system proxy. To mitigate this, we applied VPN-based packet capture as a supplementary technique, which captures \proxyBypassBypassed{} of \proxyBypassApps{} such Apps (\proxyBypassRate{}). Nevertheless, \totalResisted{} of \mechApps{} protected Apps (\totalResistedPct{}) resisted all interception attempts, meaning our reported leakage rates represent a \textit{lower bound} on the true prevalence.


\noindent\textbf{Construct Validity.} 
Our leakage definition is scoped to credentials observable in intercepted network traffic, and does not account for server-side vulnerabilities or credentials stored insecurely on-device.  
The defense mechanism labeling was performed through independent manual inspection of intercepted traffic by two authors, with disagreements resolved through discussion. Functional relevance filtering was conducted by six undergraduate evaluators using standardized exclusion criteria and majority-vote adjudication. While these manual processes may introduce classification noise, our inter-rater agreement and multi-evaluator design mitigate systematic bias.

\noindent\textbf{External Validity.}
Our dataset comprises \totalTestApps{} LLM-integrated iOS Apps collected from the US App Store in Oct 2025. This represents a temporal snapshot; Apps may have updated their security practices since data collection. 
However, the dataset spans 13 app categories with varying download volumes, suggesting that our findings are not artifacts of a particular app segment. 
Additionally, the keyword-based search strategy may miss Apps that integrate LLM functionality without using identifiable keywords in their metadata. Lastly, our findings are specific to the iOS ecosystem and the US App Store region. Leakage patterns may differ on Android or in other regional App Stores, which have different developer demographics and regulatory environments.

%% file: Conclusion.tex
\section{Conclusion}

This paper presents the first empirical study of LLM API credential leakage in iOS applications. Using \tool, a dynamic analysis framework based on runtime traffic interception, we found that \testedExploitRate{} of \totalTestApps{} LLM-integrated applications expose credentials in network traffic, of which \leakedFullPct{} are confirmed fully exploitable. Our study reveals two key insights. First, backend proxies alone do not prevent leakage. Although the proxy architecture is the recommended approach for protecting LLM API keys, JWT-based leakage through proxies remains the most prevalent pattern (\jwtPct{}).
This indicates that developers widely adopt the proxy pattern yet fail to enforce authentication and access control on the proxy itself. 
Second, the low remediation rate reflects the complexity of mitigating this vulnerability. Persistent cases mainly involve unauthenticated proxies or flawed token management, indicating that pattern-specific remediation guidance is needed alongside responsible disclosure. These findings reveal that LLM credential security is a shared responsibility across multiple stakeholders.

In future work, we plan to collaborate with LLM providers on server-side anomaly detection for compromised credentials, and to develop automated remediation tooling that generates secure proxy scaffolding with proper authentication.